# Influence of excitation frequency on Raman modes of In$_{1-x}$Ga$_x$N thin films

A. Dixit[1,*], J. S. Thakur[2], V. M. Naik[3], R. Naik[2]

[1]Center of Excellence in Energy & ICT, Indian Institute of Technology Jodhpur, Rajasthan, India 342011

[2]Department of Physics and Astronomy, Wayne State University, Detroit, MI, USA 48202

**[3]**Department of Natural Sciences, University of Michigan-Dearborn, Dearborn, MI, USA 48128

[*]ambesh@iitj.ac.in

**Abstract:**

Low energy optical modes of MBE-grown In$_{1-x}$Ga$_x$N thin films with different values of x are investigated using Raman spectroscopy. We also studied the influence of Raman excitation frequency using red and green lasers on scattering intensity of various Raman modes. For those In$_{1-x}$Ga$_x$N alloys whose bandgap energy is close to the red laser, a huge enhancement in the intensities of A$_1$(LO) mode and its 2A$_1$(LO) replica is observed when excited with red laser as compared to the green laser excitation. We found that the energies of longitudinal optical modes (A1 (LO) and 2A$_1$ (LO)) vary nonlinearly unlike the E$_2$ mode with increasing Ga atomic fraction. A Raman mode ~ 540 cm$^{-1}$ was observed in all In$_{1-x}$Ga$_x$N films with low energy red laser excitation but was absent with green laser excitation. We attribute this mode to A$_1$(TO) mode of the underneath GaN buffer layer.

**1. Introduction**

Currently various optical properties of In$_{1-x}$Ga$_x$N alloys are extensively studied because of their potential applications in full-solar-spectrum photovoltaics [1, 2, 3], high-performance light-emitting and laser diodes [4] and, solid-state lighting [5]. In addition to the investigation of their bulk properties, the quantum well structures of these alloys are also currently investigated for a variety of applications including laser diodes and μ-photoluminescence spectroscopy with InGaN quantum disks [6]. High efficiency quantum well structure light-emitting-diodes in the blue-green region of the spectrum and laser diodes emitting violet light have already been made from group-III nitride semiconductors [7] and are in commercial use. In all these devices In$_{1-x}$Ga$_x$N forms an active layer. The bandgap of these alloys can also be tuned from near-infrared (IR) to ultraviolet (UV) regimes and in addition these alloys can be fabricated to further increase their bandgap energy by quantum confinement effects [8].

Photoluminescence and other optical properties are strongly influenced by electron relaxation rates due to electron-electron and electron-phonon scattering [9, 10]. So it is important to investigate their various phonon modes and how their spectral features can be influenced by the source excitation frequency. We studied the phonon modes of $In_{1-x}Ga_xN$ alloys using Raman spectroscopy and variations in their energies with increasing Ga atomic fraction in the InN compound matrices. We also investigated the effects of Raman excitation frequency on the scattering intensity of these modes for various values of x.

## 2. Experiment

$In_{1-x}Ga_xN$ thin films with x = 0, 0.15, 0.30 and 0.54 were prepared by MBE [11, 12] technique. Thin buffer layers of AlN and GaN with thickness ~ 10 nm and 220 nm were grown to reduce the lattice mismatch between $In_{1-x}Ga_xN$ thin films and c-sapphire substrate. The thickness of the $In_{1-x}Ga_xN$ layer is ~ 0.6 μm. The carrier concentrations and Hall mobilities of these thin films were determined at room temperature using Van der Pauw method and their values are listed in Table I. We noticed a small increase in carrier concentration and a substantial decrease in mobility with increasing Ga atomic fraction in the $In_{1-x}Ga_xN$ films. The well-defined interference fringe width from UV-Vis-NIR reflection spectrum was used to estimate the thickness of these thin films whose values are given in Table (1).

## 3. Results and Discussion

A strong electron-phonon Fröhlich interaction in wurtzite crystals can lead to a large enhancement in light scattering by polar LO phonons when an incident photon's energy is close to band gap energy of semiconductor [13, 14]. In many cases it has also led to excitation of replica phonon modes. In wurtzite $In_{1-x}Ga_xN$ structure, there are six Raman active modes--$E_2$(low), $E_2$(high), $A_1$(TO), $A_1$(LO), $E_1$(TO), and $E_1$(LO) [15, 16] which are observed in the first order Raman spectrum of $In_{1-x}Ga_xN$ films. Other optical modes --$B_1$(low) and $B_1$(high) -- are silent in wurtzite nitrides but are occasionally observed in disordered films [17]. In this paper, we investigated the role of excitation energy on the Raman intensity of these modes by choosing two laser energies— the green laser (514.5 nm = 2.41 eV) having energy larger than the bandgap energy of all the $In_{1-x}Ga_xN$ films we have studied here and the red laser (785 nm = 1.58 eV) having energy between the band gap energy of x = 0.3 film ($E_g$ = 1.30 eV) and of x = 0.54 film ($E_g$ = 1.85 eV). Out of the six Raman active modes, we observed $E_2$(high) and the first-order $A_1$(LO) Raman modes when the samples were excited by red laser (Fig. 1). Along with these modes, $2A_1$(LO) replica mode is also observed in all the samples. In between $E_2$(High) and $A_1$(LO) modes, one can also see a mode around 540 cm$^{-1}$ in all the $In_{1-x}Ga_xN$ films. A similar mode has been observed [18] in post annealed InN thin films and has been marked as an unidentified feature. However, we associate this mode with $A_1$(TO) mode of GaN originating from the GaN buffer layer used in all the films. This mode is not observed with green laser (Fig. 2(a)) because of the stronger absorption of the green laser by the top $In_{1-x}Ga_xN$ layer. As Ga atomic fraction in $In_{1-x}Ga_xN$ is increased, the energies of these modes ($E_2$(High), $A_1$(LO), and $2A_1$(LO)) increase as shown in Fig.3(a), as expected. However there is a dramatic decrease in the intensity of $E_2$(high) mode relative to $A_1$(LO) mode accompanied by increase in damping due to increase in the lattice disorder from Ga atoms substitution. At x = 0.54, $E_2$(high) mode has almost disappeared into the noise. With increasing x value, the absorption of the laser energy decreases in the $In_{1-x}Ga_xN$ layer due to increase in the bandgap values and at x = 0.54, the bandgap value ($E_g$ = 1.86 eV) of $In_{1-x}Ga_xN$ layer becomes larger than the laser energy. This leads to a stronger transmission of the laser energy into the subsequent layers—GaN and Sapphire layers-- and as a result intense excitations of many sapphire modes as marked by circles along with the $A_1$(TO) of GaN marked as * (top panel, x = 0.54 in Fig. 1). We also observed an intense structure at 1070 $cm^{-1}$. Similar structure, however with less intensity, has been associated with the combination of $E_g$(432 $cm^{-1}$) GaN mode and $A_{1g}$ (645 $cm^{-1}$) sapphire mode[19]. Because of stronger excitation of the sapphire modes in this film, the observed mode could be a result of this combination.

In Fig. 2(a) we show the Raman spectra of $In_{1-x}Ga_xN$ films when excited off-resonance with green laser. A big change in the Raman intensities of these modes is observed when compared with the corresponding Raman spectral intensities obtained using red laser. The green laser energy is larger than the bandgap energies of the $In_{1-x}Ga_xN$ films, so there will be a negligible transmission of green laser energy into the GaN buffer layer underneath the $In_{1-x}Ga_xN$ films and sapphire substrate. As a result, no sapphire and GaN modes are observed in $In_{1-x}Ga_xN$ thin films with green laser, as shown in Fig 2(a). With increasing atomic fraction of Ga, the bandgap energy of $In_{1-x}Ga_xN$ layer increases and begins to tune with the laser energy. This increases the Fröhlich electron-phonon interaction leading to a stronger excitation of $A_1$(LO) and $2A_1$(LO) phonons. At x = 0.54, the $A_1$(LO) and $2A_1$(LO) phonons become quite intense compare to the $E_2$ modes. On the contrary, the Raman intensities of $E_2$(low) and $E_2$(high) modes decrease with increasing values of x and at x = 0.54, these modes cease to exist due to overdamping by the substitutional disorder of Ga atoms. Note that $E_2$(low) mode could not be observed when excited with red laser. In Fig. 2(b), we also show the shift in the energy of $E_2$(low) mode with increasing values of x. The energies of the observed $A_1$(LO) and $2A_1$(LO) modes increase nonlinearly (Fig. 3(a)) while those of $E_2$ modes vary almost linearly with increasing Ga atoms concentration. We attribute this nonlinear behavior to the c-axial strain in our films due to their small thickness affecting lattice polarization along the c-axis more than in other directions [20]. Similar behavior of these modes is also observed for the red laser when Ga concentration is increased. Fig. 3(b) shows variation in the relative intensities of $A_1$(LO) and $E_2$(High) modes for red and green lasers. Clearly the ratio $A_1$(LO) /$E_2$(High) is quite large for red laser and varies quite strongly as compared to that of the green laser with increasing concentration of Ga. At a particular value of x, there is no change in the frequency of these modes when energy of the laser is changed, however, the intensities of some of the modes change dramatically.

## 4. Summary

In summary, we have investigated the intensity of Raman scattering in $In_{1-x}Ga_xN$ thin films when their bandgap energy is close to the excitation frequency of the laser. We found that tuning of electronic bandgap transition of the films with the laser excitation frequency dramatically influences the scattering intensities of $A_1$(LO), $2A_1$(LO), $E_2$(Low) and, $E_2$(high) modes.

**Acknowledgements:**

Authors are thankful to Dr. W. J. Schaff for providing us the $In_{1-x}Ga_xN$ samples.

Table I. Room temperature electronic and optical properties of $In_{1-x}Ga_xN$ thin films

| Sample | Thickness ($\mu m$) | Band gap (eV) | Carrier concentration ($cm^{-3}$) | Mobility ($cm^2/V.S$) |
|---|---|---|---|---|
| GS1814 InN | 0.6 | 0.77 | $1.3 \times 10^{18}$ | 900 |
| GS1651 $In_{0.85}Ga_{0.15}N$ | 0.5 | 0.96 | $1.43 \times 10^{18}$ | 700 |
| GS1942 $In_{0.70}Ga_{0.30}N$ | 0.5 | 1.27 | $2.3 \times 10^{18}$ | 140 |
| GS1651 $In_{0.46}Ga_{0.54}N$ | 0.4 | 1.85 | $1.1 \times 10^{18}$ | 20 |

Fig. 1. Raman spectra of $In_{1-x}Ga_xN$ (x = 0, 0.15, 0.30 and 0.54) thin films using red laser excitation source (785 nm = 1.58 eV). The dashed lines are guides to eyes tracing variation of $A_1(LO)$ and its replica $2A_1(LO)$ mode with increasing Ga atomic fraction. *(red) indicates the $A_1(TO)$ GaN mode and circles show the sapphire modes observed in $In_{0.46}Ga_{0.54}N$ thin films

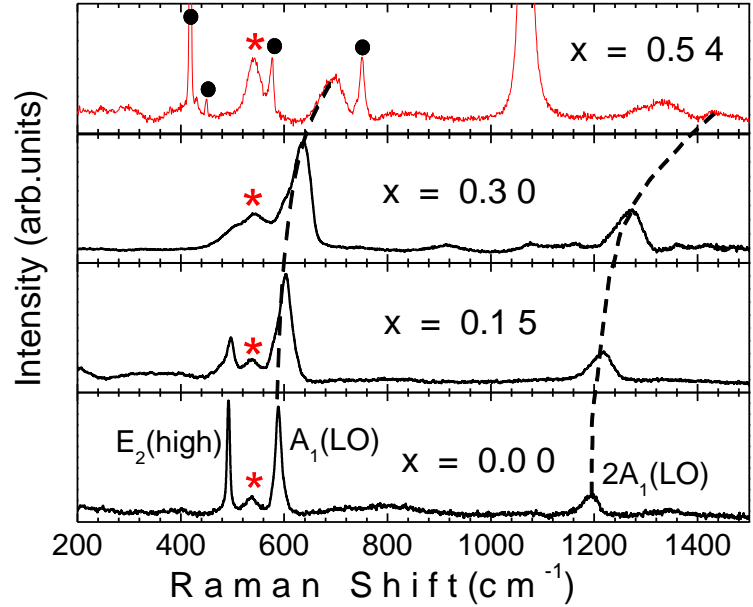

Fig. 2(a) Raman spectra of $In_{1-x}Ga_xN$ (x = 0, 0.15, 0.30 and 0.54) thin films and 2(b) variation in the energy of $E_2(Low)$ using green laser excitation source (514.5 nm = 2.41 eV). The dashed lines are guides to eyes tracing variations of $E_2(low)$, $A_1(LO)$ and its replica $2A_1(LO)$ mode energies with increasing Ga atomic fraction.

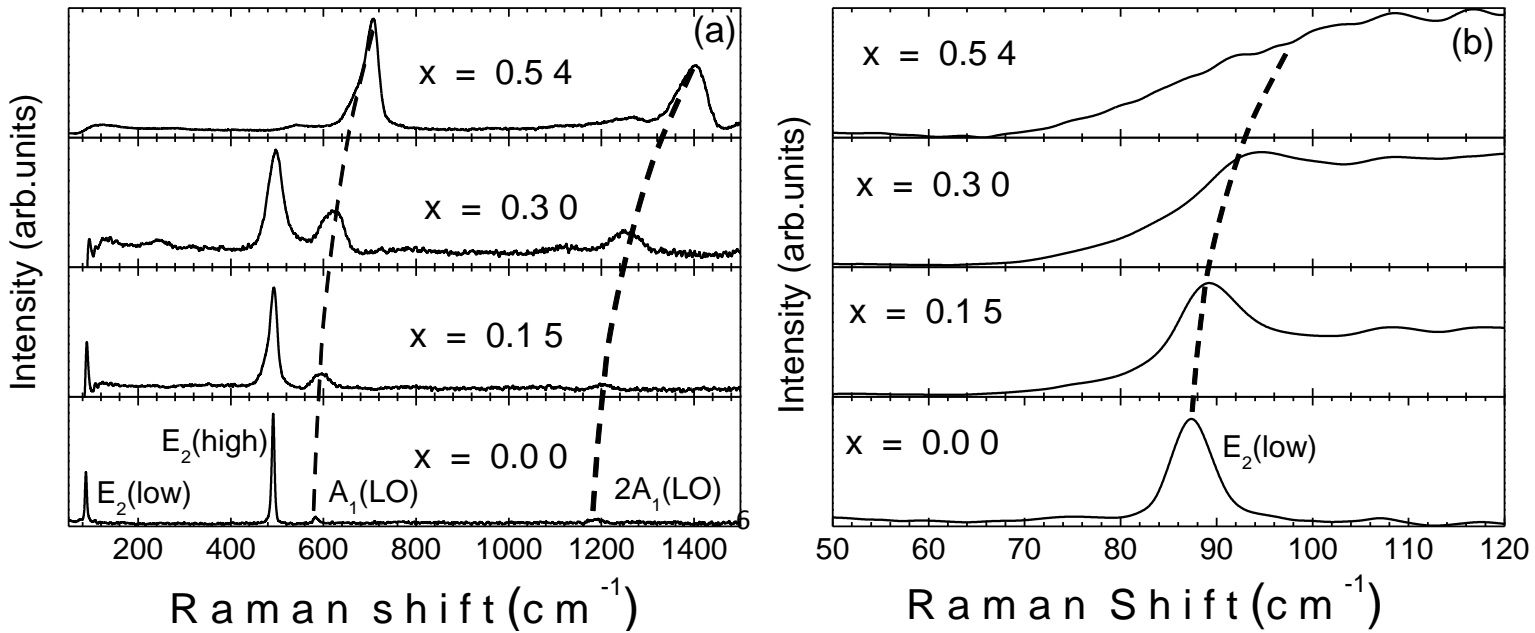

Fig. 3(a) Energy variations of $1A_1(LO)$, $2A_1(LO)$, $E_2(Low)$ and $E_2(High)$ modes with respect to Ga atomic fraction in $In_{1-x}Ga_xN$ alloy thin films

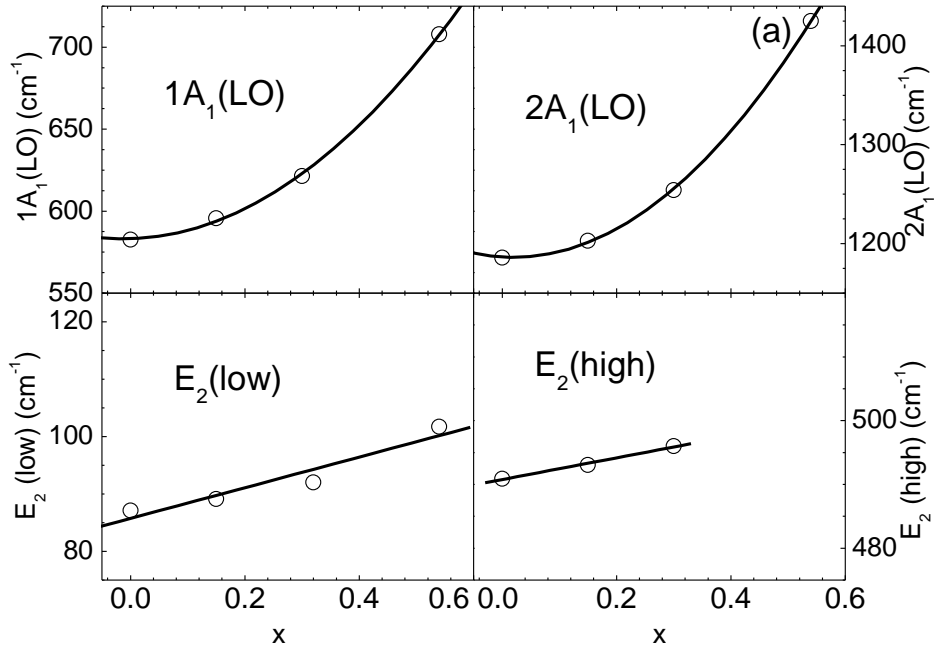

Fig. 3(b) Variation in the ratio of $A_1(LO)/E_2(High)$ with respect to Ga atomic fraction for $In_{1-x}Ga_xN$ alloy thin films for red and green lasers.

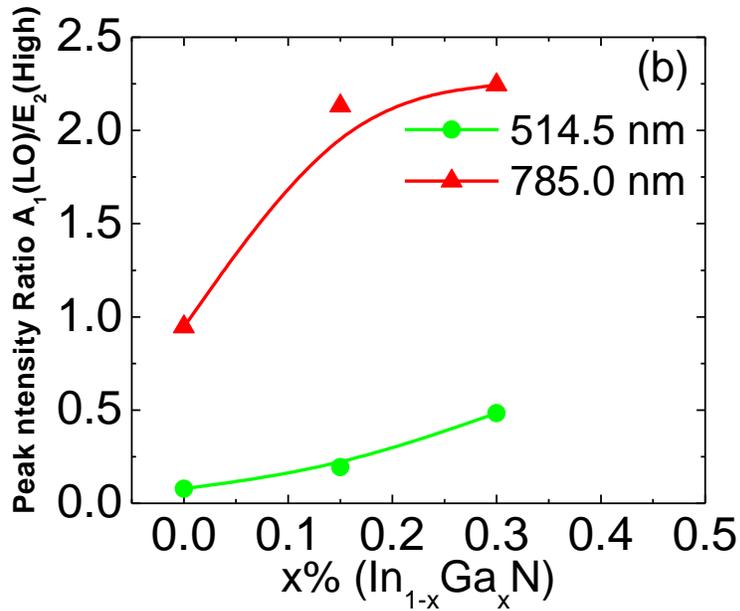